\documentstyle[12pt,psfig]{article}
\begin{document}
\begin{center}
\Large{\bf A Second Look at Single Photon Production in $S+Au$ Collisions
 at 200 A$\cdot$GeV and Implications for Quark Hadron Phase Transition}
\vskip 0.2in

\large{Dinesh Kumar Srivastava$^1$  and Bikash Chandra Sinha$^{1,2}$}
\vskip 0.2in

\large{\em $^1$ Variable Energy Cyclotron Centre,\\
 1/AF Bidhan Nagar, Calcutta 700 064, India\\

$^2$ Saha Institute of Nuclear Physics,\\
1/AF Bidhan Nagar, Calcutta 700 064, India\\}

\vskip 0.2in

Abstract

\vskip 0.2in

\end{center}

We reanalyze the production of single photons in $S+Au$ collisions
at CERN SPS to investigate: i) the consequences of using a 
much richer equation of state for hadrons than the one used in an earlier
study by us; and, ii) to see if the recent estimates of photon
production in quark-matter (at two loop level) by Aurenche et al.
are consistent with the upper limit of the photon production
measured by the WA80 experiment. We find that the measured upper limit
is consistent with a quark hadron phase transition. The measured 
upper limit 
is also consistent with a scenario where no phase transition
takes place, but where the hadronic matter reaches a density
of several hadrons per unit volume; which is rather
unphysical. 

\newpage

The publication of the upper limit of the production of
single photons in $S+Au$ collisions at CERN SPS~\cite{wa80}
by the WA80 experiment has been preceded and followed by several papers
exploring their connection to the so-called quark-hadron phase
transition. Thus an early work by the present authors~\cite{prl}, argued
that the measurement {\em is} consistent with a scenario where a quark-gluon
plasma is formed at some time $\tau_0\approx$ 1 fm/$c$, which expands
and cools, converts into a mixed phase of quarks, gluons, and hadrons,
and ultimately undergoes a freeze-out  from a state of hadronic gas
consisting of $\pi$, $\rho$, $\omega$, and $\eta$ mesons. 
On the other hand, when the initial state is assumed to consist
of (the same) hadrons, the resulting large initial temperature leads to a
much larger production of single photons, in a gross
disagreement with the data.

Since then, several authors have looked at the production of single
photons in these collisions, using varying evolution scenarios, and
including the effects of changing (baryon) density and temperature
on the rate of production of photons from the hadronic matter.

Thus for example, Cleymans, Redlich, and Srivastava~\cite{jean}
used a hadronic equation of state which included {\em {all}} hadrons having
a mass of up to 2.5 GeV, from the particle data book in a complete
thermal and chemical equilibrium. In this approach, the production
of photons in the phase-transition and the no-phase transition scenarios
(for $Pb+Pb$ collisions at CERN SPS) was predicted to be quite 
similar. However, the authors  also noted that, the no-phase
transition scenario necessitated a hadronic matter, where 2--3 hadrons
had to be accommodated within a volume of $\approx$ 1 fm$^3$, where the hadronic
picture should surely break-down (see also~\cite{vesa}).

All the above studies were performed using the (one-loop) evaluation
of single photons from the quark matter~\cite{joe,rolf} and hadronic reactions 
using varying effective Lagrangians. An attempt was also made to evaluate
the single photons from the quark matter due the bremsstrahlung
processes within the soft-photon approximation which were found to
make a large contribution at smaller $p_T$~\cite{kryz,dipali}.

A new dimension (and hope) has recently been added to these efforts
by the evaluation of the rate of single photon production from the
quark matter to the order of two-loops by Aurenche et al~\cite{pat}. 
The two most interesting results are: (i) the dominance of the bremsstrahlung
process for all momenta over the Compton plus annihilation contributions
included in the one-loop calculations mentioned above, and even more
importantly; (ii) a very large contribution by a new mechanism
which corresponds to the annihilation of a off-shell quark ( produced
in a scattering with  a
quark or a gluon) by an anti-quark. This considerably enhances the
production of single photons at SPS, RHIC, and LHC energies and in
a very significant departure from the (usual and some-what pessimistic) belief,
the dominant number of photons are now predicted to have their origin
in the quark matter~\cite{dks}, if the initial state could be approximated
 as an equilibrated plasma. While the results for a (chemically)
non-equilibrated plasma are awaited, a large production of high
$p_T$ photons has been predicted from the pre-equilibrium stage of the
parton cascade model~\cite{pcmphot}.
 Taken in its entirety, it presents a very
positive development for the expected observability  of single 
photons from relativistic heavy ion collisions.

However, this also raises an important issue.  The 
analysis of Ref.~\cite{prl} has to be repeated to see if the
newly identified processes contributing to the single photons from
the quark matter remain consistent with the upper limit of the WA80
experiment. In case it overshoots the upper limit, we have to identify the
reasons for it.

Let us briefly recall the sources of single photons from the quark
matter.  During the QGP phase, the single photons
originate from Compton 
($q\,(\overline{q})\,g\,\rightarrow\,q\,(\overline{q})\,\gamma$)
and annihilation ($q\,\overline{q}\,\rightarrow\,g\,\gamma$)
processes~\cite{joe,rolf} as well as bremsstrahlung processes
($q\,q\,(g)\,\rightarrow\,q\,q\,(g)\,\gamma$)~\cite{kryz,dipali}. During
the pre-quilibrium phase, which can be treated within the
parton cascade model~\cite{pcm},
the fragmentation of time-like 
quarks ($q\,\rightarrow\,q\,\gamma$) produced
in (semi)hard multiple scatterings  
leads to a substantial production of photons (flash of
photons!), whose $p_T$ is decided by the $Q^2$ of the scatterings and not the
temperature as in the calculations mentioned earlier~\cite{pcmphot}.
Of course, we must now admit the suggestion of
Aurenche et al~\cite{pat} that the
production of photons in a QGP  evaluated up to two loops, leads
to a large bremsstrahlung contribution
(see Ref.~\cite{kryz,dipali} for early estimates
within a soft photon approximation) as well as the 
new mechanism for the production of hard photons through the annihilation
of quarks with scattering,
which completely dominates the emission of hard photons. 

The rate for the production of hard photons evaluated to one
loop order using the effective theory based on resummation of
hard thermal loops is given by~\cite{joe,rolf}:
\begin{equation}
E\frac{dN}{d^4x\,d^3k}=\frac{1}{2\pi^2}\,\alpha\alpha_s\,
                          \left(\sum_f e_f^2\right)\, T^2\,
                       e^{-E/T}\,\ln(\frac {cE}{\alpha_s T})
\end{equation}
where the constant $c\approx$ 0.23.  The summation runs over the
 flavours of the quarks and $e_f$ is the electric charge of the
quarks in units of charge of the electron. The rate of production
of photons due to the bremsstrahlung processes evaluated by
Aurenche et al is given by:
\begin{equation}
E\frac{dN}{d^4x\,d^3k}=\frac{8}{\pi^5}\,\alpha\alpha_s\,
                          \left(\sum_f e_f^2\right)\, 
                        \frac{T^4}{E^2}\,
                       e^{-E/T}\,(J_T-J_L)\,I(E,T)
\end{equation}
where $J_T\approx$ 4.45 and $J_L\approx - $4.26 for 2 flavours and 3
colour of quarks. $I(E,T)$ stands for;
\begin{eqnarray}
I(E,T)&=&\left[ 3\zeta(3)+\frac{\pi^2}{6}\frac{E}{T}+
        \left(\frac{E}{T}\right)^2\ln(2)\right.\nonumber\\
        & &+4\,Li_3(-e^{-|E|/T})+2\,
            \left(\frac{E}{T}\right)\,Li_2(-e^{-|E|/T})\nonumber\\
        & &\left. -\left(\frac{E}{T}\right)^2\,\ln(1+e^{-|E|/T})\right]~,
\end{eqnarray}
and the poly-logarith functions $Li$ are given by;
\begin{equation}
Li_a(z)=\sum_{n=1}^{+\infty}\frac{z^n}{n^a}~~.
\end{equation}

 And finally the contribution of the $q\overline{q}$ annihilation
with scattering obtained by them is given by:
\begin{equation}
E\frac{dN}{d^4x\,d^3k}=\frac{8}{3\pi^5}\,\alpha\alpha_s\,
                          \left(\sum_f e_f^2\right)\, ET \,
                          e^{-E/T}\,(J_T-J_L)
\end{equation}

We assume that a chemically and thermally equilibrated quark-gluon
plasma is produced in such collisions at the time $\tau_0=$ 1 fm/$c$,
and that one could use
 the Bjorken condition~\cite{bj};
\begin{equation}
\frac{2\pi^4}{45\zeta(3)}\,\frac{1}{\pi R_T^2}\frac{dN}{dy}=4 a
T_0^3\tau_0
\end{equation}
to obtain an estimate of the initial temperature. We have
chosen the particle rapidity density as 225 for the $S+Au$ collision
at the CERN SPS energy, with the transverse dimension decided by the
radius of the $S$ nucleus, and taken $a=37\pi^2/90$ for
a plasma of mass-less quarks (u and d) and gluons. In the case of no 
phase transition we estimate the temperature to yield the same entropy
as in the above~\cite{jean}.

We assume the phase transition to take place at $T=$ 160 MeV, and the
freeze-out to take place at 120 MeV.
We use a hadronic equation of state consisting of {\em {all}} the hadrons and
resonances from the particle data table which have a mass less then 2.5
GeV~\cite{jean}. The rates for the hadronic matter have been 
obtained~\cite{joe}
from a two loop approximation of the photon self energy 
using a model where $\pi-\rho$ interactions have been included. The 
contribution of the $A_1$ resonance is also included according to the
suggestions of Xiong et al~\cite{li}. The relevant hydrodynamic equations are
solved using the procedure~\cite{hydro} discussed earlier and
an integration over history of evolution is performed~\cite{jean}. 

In Fig.~1 we show our results for the phase transition scenario.
As remarked in the figure caption there, the dot-dashed curve gives the
contribution of the quark-matter evaluated to the order of one loop,
 the dashed curve gives the
contribution of the hadronic matter, and the solid curve gives the
sum of the two. We have also separately given the 
{\em pre-equilibrium} contribution
evaluated within a parton cascade model~\cite{pcmphot}. This is estimated by
normalizing the corresponding predictions for
$S+S$ collision at the SPS energy 
with the ratio of the nuclear thicknesses $T_{SAu}/T_{SS}$ at 
$b=0$. It is interesting  to see that the non-exponential component
apparent in the measured upper limit can be identified with the
 pre-equilibrium contribution.

It is seen that the photon yield stays below the upper limit at all
$p_T$, and most significantly, the dominant contribution is
from the radiation from the hadronic matter.

The corresponding results with rates evaluated to the order of two-loops
are given in Fig.~2 using similar notations.
We now see that the evaluated photon yield has a dominant
contribution from the quark-matter, as remarked earlier~\cite{dks}.
 We also note that
the predicted yield closely follows the shape of the
measured upper limit over the
entire range of $p_T$.

We see that the evaluated photon yield exhausts the upper limit at all
$p_T$. However, considering that the measurements represent an upper limit,
it still leaves a scope for a discussion of scenarios which may reduce the
yield of single photons. The fore-most consideration, and which
is also most likely, would be an initial state where
 the quark-gluon plasma is {\em not} in chemical
equilibrium. In fact several studies at RHIC and LHC energies
do suggest that the initial state of the plasma may not be in a 
state of chemical equilibrium~\cite{pcm}, though thermal equilibrium may
be attained rather quickly~\cite{therm}.

In Fig.~3 we have shown our predictions for the scenario when
no phase transition takes place.  We again see that at least beyond
$p_T$ equal to 1 GeV/$c$, the  estimated single photon yield 
is consistent with the upper limit, though it is smaller than the
upper limit both at the lower and the upper end of the $p_T$
spectrum. However, we have to emphasize that this description
involves a hadronic gas which has a number density of several
hadrons/fm$^3$, which is rather un-physical, and 
we have reservations about this 
description. We reiterate that the picture leading to
the Fig.~2 (or 1) is more likely. We may also add that for such high
hadronic densities almost all prescriptions for accommodating
finite size effects in the hadronic equation of state will either
break-down or imply a very high energy density to overcome the
so-called hard-core repulsion of the hadrons at very short
densities.

Before concluding, it is of interest to add a comment on the shape of the
predicted spectra (Figs.~1--3) in comparison to the measured
upper limit. We have already
remarked that the predictions lie considerably below the upper
limit in Figs. 1 and
3, at lower $p_T$. Recall that the hadronic reactions considered in the
study include the process $\pi \pi \rightarrow \rho \gamma$ which
is known to be equivalent to the bremsstrahlung process $\pi \pi \rightarrow
\pi \pi \gamma$ (see Ref.~\cite{kryz,dipali}). Thus we realize that
the bremsstrahlung process in the quark matter contained in the two-loop
evaluations of Aurenche et al~\cite{pat} plays an important role
in getting the right shape of the spectrum at lower $p_T$. We do not have
to repeat that the pre-equilibrium contribution leads to the right shape
at higher $p_T$.  

We conclude that the newly obtained rates for emission of
photons from QGP (evaluated to the order of two loops), which are
much larger than the corresponding results for the one-loop
estimates yield single photons which are in agreement with the
upper limit of the data obtained by the WA80 experiment for the
$S+Au$ collisions at CERN SPS, and support a description
where a quark-gluon plasma is formed. 

We add that considering that the data represent the
upper limit we can, in principle, admit a scenario  which has a chemically
non-equilibrated plasma at the time $\tau_0$, and which
will lead to a smaller radiation of photons from the quark phase.

\section*{Acknowledgments} 
One of us (dks) gratefully acknowledges the hospitality of University of
Bielefeld where part of this work was done.  He would also like to thank
 Terry Awes and Axel Drees whose questions led us to take 
a second look at the measurements discussed here in the light of the 
recent works of Ref.~\cite{jean,pat}.

\newpage

%%%%%%%%%%%%%%%%%%% figure 1 %%%%%%%%%%%%%%%%%%%%%%%%%
\begin{figure}
\psfig{file=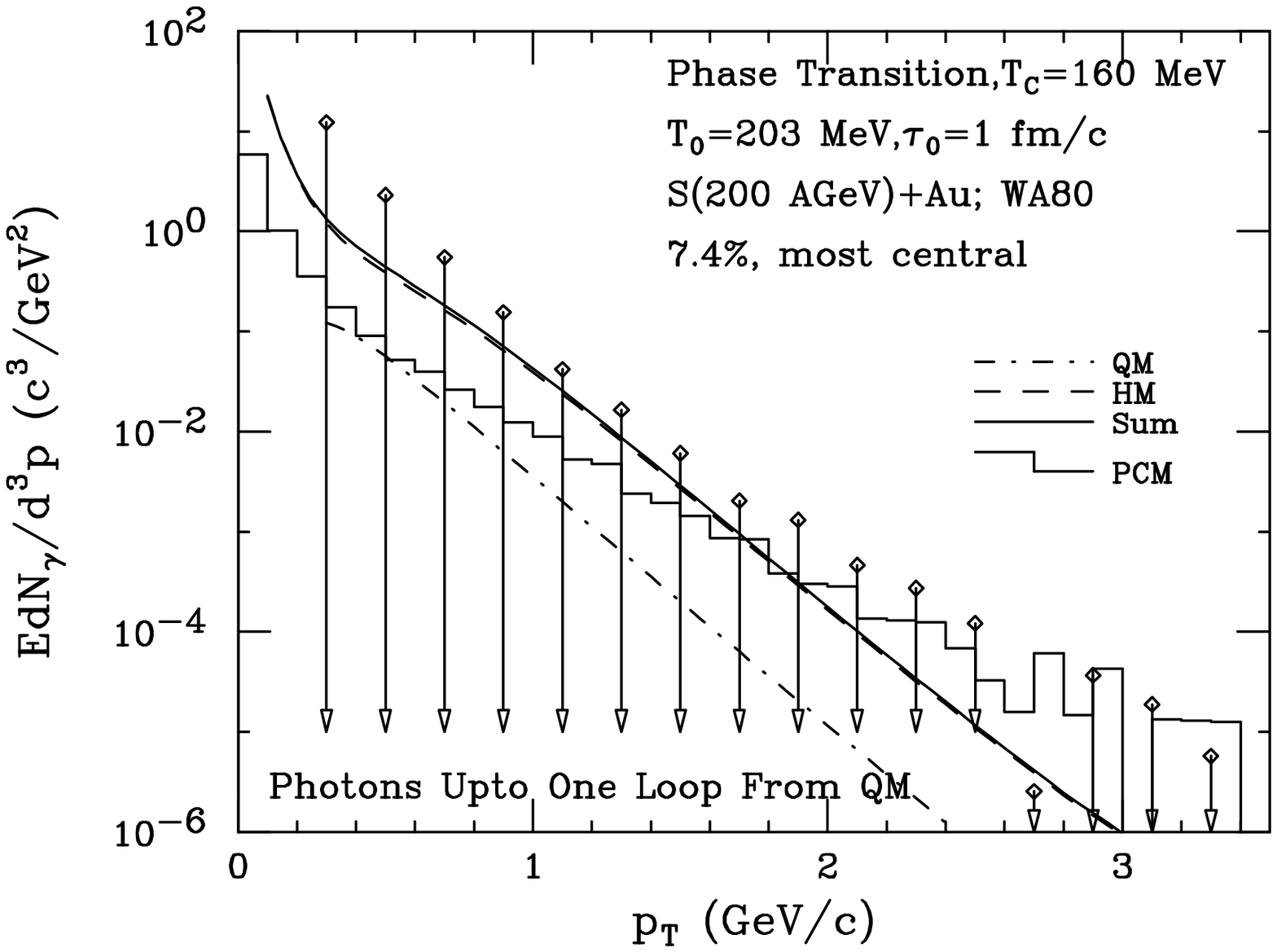,height=12cm,width=15cm}
\vskip 0.1in
\caption{ Single photon production in $S+Au$ collision at CERN SPS.
An equilibrated (chemically and thermally) quark-gluon plasma is
assumed to be formed at $\tau_0$ which expands, cools, gets into a
mixed phase and undergoes freeze-out. QM stands for radiations from the
quark matter in the QGP phase and the mixed phase. HM, likewise denotes
the radiation from the hadronic matter in the mixed phase and the
hadronic phase and Sum denotes the sum  of the contributions from the
equilibrium phase. The histogram shows the pre-equilibrium contribution
evaluated in a parton cascade model. The radiations from the 
quark-matter are evaluated to the order of one-loop.
}
\end{figure}
%%%%%%%%%%%%%%%%%%% end of figure 1%%%%%%%%%%%%%%%%%%%%%%%%%

\newpage

%%%%%%%%%%%%%%%%%%% figure 2 %%%%%%%%%%%%%%%%%%%%%%%%%
\begin{figure}
\psfig{file=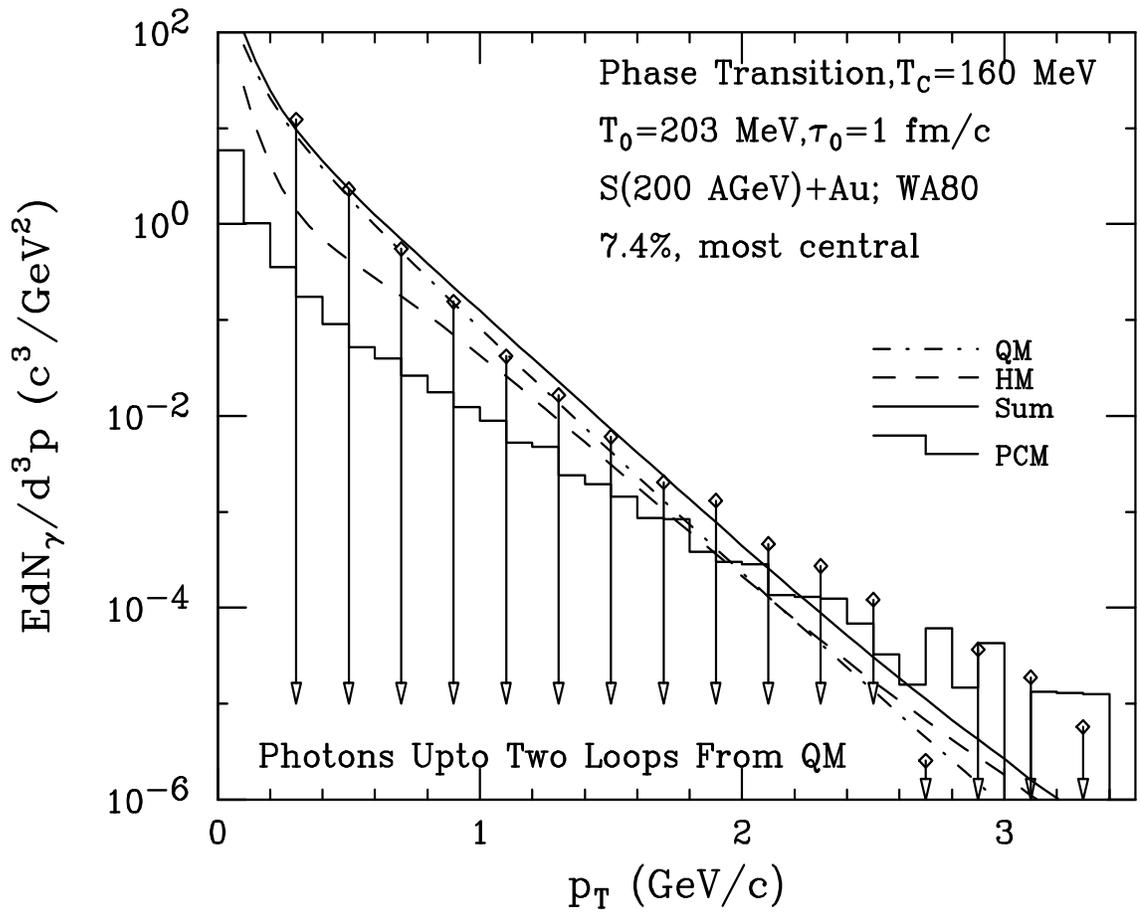,height=12cm,width=15cm}
\vskip 0.1in
\caption{ Same as Fig.~1, with the radiations from the
quark-matter evaluated to the order of two loops.
}
\end{figure}
%%%%%%%%%%%%%%%%%%% end of figure 2%%%%%%%%%%%%%%%%%%%%%%%%%

\newpage
%%%%%%%%%%%%%%%%%%% figure 3 %%%%%%%%%%%%%%%%%%%%%%%%%
\begin{figure}
\psfig{file=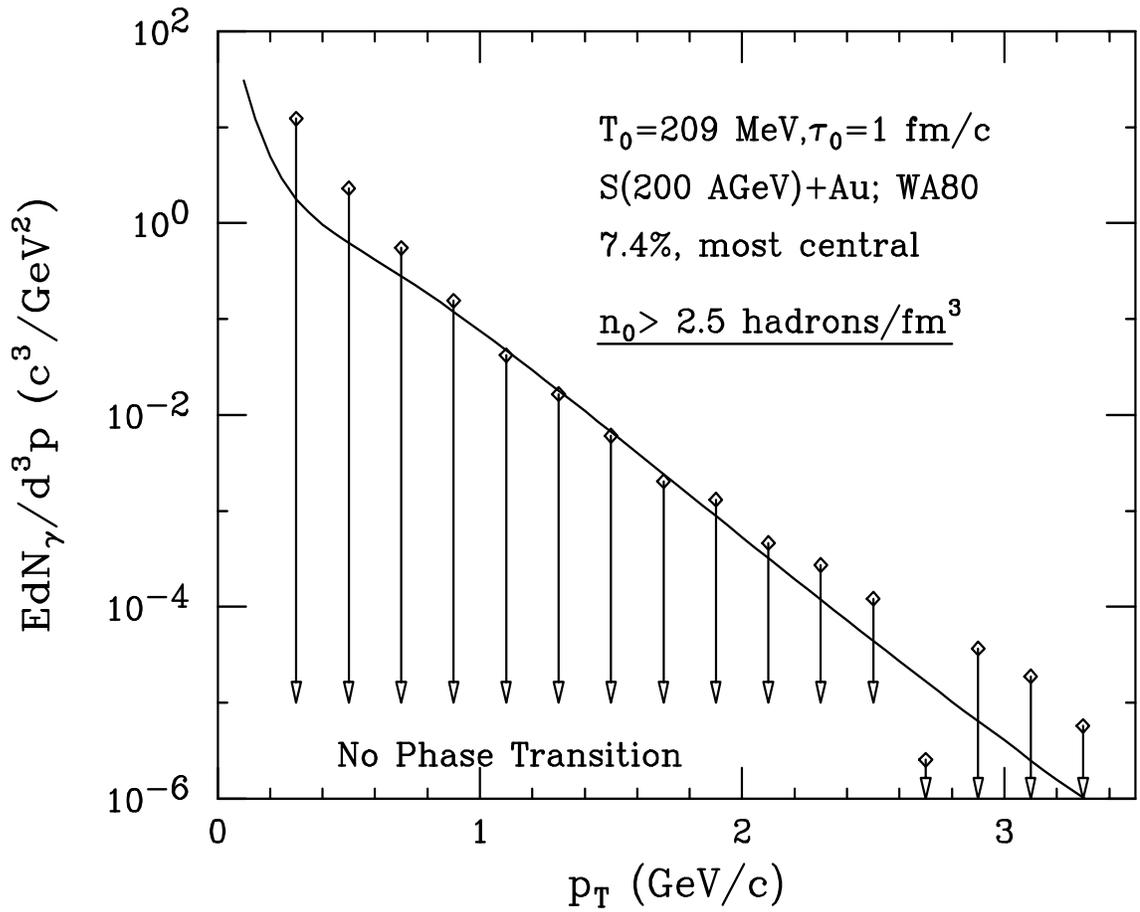,height=12cm,width=15cm}
\vskip 0.1in
\caption{ Same as Fig.~1, but without a phase transition; i.e.,
a hot hadronic gas is assumed to be formed at $\tau_0$. Note, however,
in this case the initial density of hadrons exceeds several
hadrons/fm$^3$.
}
\end{figure}
%%%%%%%%%%%%%%%%%%% end of figure 3%%%%%%%%%%%%%%%%%%%%%%%%%
\end{document}